\documentclass[11pt,titlepage]{article}
\usepackage{graphicx}
\usepackage{rotating}
\usepackage{bm}
\usepackage{amsmath}
\usepackage{amssymb}
\usepackage{hyperref}
\usepackage{setspace}
\usepackage{xr}
\usepackage[round,numbers,sort&compress]{natbib}
\usepackage[margin=1in]{geometry}

\def\kon{k_{\rm on}}
\def\konc{k_{\rm on}c}
\def\koff{k_{\rm off}}
\def\Kon{K_{\rm on}}
\def\Konc{K_{\rm on}c}
\def\Koff{K_{\rm off}}
\def\mum{$\mu$m}
\def\pers{ s$^{-1}$}

\def\pernm{ nM$^{-1}$}

\title{Motor protein accumulation on antiparallel microtubule
  overlaps}

\author{H.-S. Kuan and M. D. Betterton\thanks{ Corresponding author.
    Address: Department of Physics, University of Colorado Boulder 390
    UCB Boulder, CO~803095, U.S.A.,
    Tel.:~(303)735-6135 }\\
  Department of Physics, \\
  University of Colorado Boulder, Boulder, CO }

\date{}

\begin{document} 
\maketitle

\abstract{Biopolymers serve as one-dimensional tracks on which motor
  proteins move to perform their biological roles. Motor protein
  phenomena have inspired theoretical models of one-dimensional
  transport, crowding, and jamming. Experiments studying the motion of
  Xklp1 motors on reconstituted antiparallel microtubule overlaps
  demonstrated that motors recruited to the overlap walk toward the
  plus end of individual microtubules and frequently switch between
  filaments. We study a model of this system that couples the totally
  asymmetric simple exclusion process (TASEP) for motor motion with
  switches between antiparallel filaments and binding kinetics.  We
  determine steady-state motor density profiles for fixed-length
  overlaps using exact and approximate solutions of the continuum
  differential equations and compare to kinetic Monte Carlo
  simulations.  Overlap motor density profiles and motor trajectories
  resemble experimental measurements. The phase diagram of the model
  is similar to the single-filament case for low switching rate, while
  for high switching rate we find a new low density-high density-low
  density-high density phase.  The overlap center region, far from the
  overlap ends, has a constant motor density as one would na\"ively
  expect.  However, rather than following a simple binding
  equilibrium, the center motor density depends on total overlap
  length, motor speed, and motor switching rate.  The size of the
  crowded boundary layer near the overlap ends is also dependent on
  the overlap length and switching rate in addition to the motor speed
  and bulk concentration.  The antiparallel microtubule overlap
  geometry may offer a previously unrecognized mechanism for
  biological regulation of protein concentration and consequent
  activity.

  }

\clearpage

\section*{Introduction}

The motion of motor proteins on biopolymers is important for diverse
biological processes \cite{bray_cell_2000}.  Actin, microtubules, and
nucleic acids can serve as one-dimensional tracks on which motor
proteins (including myosins, kinesins, helicases, and ribosomes) move
\cite{kolomeisky_motor_2015,chowdhury_modeling_2013}. Motors must
accumulate on filaments in sufficient density to perform their
biological roles.  This motor accumulation along a biopolymer is
affected by two key effects: the directional walking of motors and
motor binding to/unbinding from the filament.
 
Motor accumulation on filaments is related to extensive theoretical
work on asymmetric exclusion processes (ASEP)
\cite{helbing_traffic_2001}. In ASEP models, particles move on a
one-dimensional (1D) lattice by biased hopping and experience
excluded-volume interactions with other particles. These models have
been applied to diverse examples of one-dimensional nonequilibrium
transport ranging from molecular motors to vehicular and pedestrian
traffic.  Active particle motion leads to non-zero flux of particle
density and nontrivial flux and density profiles.  To compare to
experimental studies of molecular motors, ASEP models have been
extended to incorporate important biophysical ingredients such as
binding kinetics.  A model extending the totally asymmetric exclusion
process (TASEP) to include motor binding and unbinding (Langmuir
kinetics) predicted motor density profiles along single fixed-length
filaments \cite{parmeggiani_totally_2004}.  Experimental work measured
kinesin-8 motor protein traffic jams on microtubules and found good
agreement with the predicted density profiles
\cite{leduc_molecular_2012}.

Motor motion on cytoskeletal filaments is important for biological
length regulation, including regulation of the length of the polymer
\cite{varga_kinesin8_2009}, lengths of overlap regions between
filaments \cite{bieling_minimal_2010}, and the length of cytoskeletal
assemblies such as the mitotic spindle
\cite{goshima_length_2005,walczak_xkcm1_1996} and even whole cells
\cite{rivero_role_1996,revenu_coworkers_2004}. The case of regulation
of microtubule (MT) length has seen the most work. MTs undergo
nonequilibrium polymerization dynamics characterized by switching
between distinct growing and shrinking states. While this dynamic
instability alone leads to a broad distribution of MT lengths
\cite{dogterom_physical_1993}, numerous proteins targeted to MT ends
modify their dynamics and can dramatically alter the length
distribution \cite{drummond_regulation_2011}. Single MTs can have
their length regulated, for example, by kinesin-8 motors that walk
with directional bias and shorten the MT from its end
\cite{gupta_endspecific_2006,varga_yeast_2006,varga_kinesin8_2009}. Theoretical
work on TASEP-like models has described how length-dependent
depolymerization affects otherwise static filaments
\cite{hough_microtubule_2009,varga_kinesin8_2009,reese_crowding_2011},
filaments with simplified polymerization kinetics,
\cite{govindan_length_2008,johann_length_2012,melbinger_microtubule_2012,reese_molecular_2014},
and dynamic MTs
\cite{tischer_providing_2010,kuan_biophysics_2013,gluncic_kinesin8_2015}.

Because the mitotic spindle includes arrays of overlapping
antiparallel MTs at the spindle midzone, regulation of MT overlaps is
important for mitotic spindle function and cytokinesis. The MT
crosslinking protein PRC1/Ase1/MAP65 and kinesin-4 motors
(chromokinesins) play roles in maintenance of the spindle midzone in
anaphase
\cite{kurasawa_essential_2004,zhu_cell_2005,khmelinskii_cdc14regulated_2007},
along with other motors and MAPs \cite{fededa_molecular_2012}. Direct
binding interactions of PRC1 and kinesin 4 can cause length-dependent
accumulation at plus ends of single MTs
\cite{subramanian_marking_2013}.  Bieling, Telley, and Surrey (BTS)
reconstituted a minimal system of stable antiparallel MT overlaps in
which PRC1 bound preferentially to overlapping regions of antiparallel
MTs \cite{bieling_minimal_2010} (see also
\citet{subramanian_insights_2010}). PRC1 recruited the kinesin-4 motor
Xklp1 to the overlap.  Xklp1 motors could bind to and unbind from the
MTs, walk toward the plus end of each MT, and switch between the two
MTs at a relatively high rate \cite{bieling_minimal_2010}. Motors
present near the MT plus ends slowed the polymerization speed,
consistent with earlier work showing that Xklp1 inhibits dynamic
instability \cite{bringmann_kinesinlike_2004} and affects spindle MT
mass \cite{castoldi_chromokinesin_2006}. As a result, antiparallel MT
overlaps reached a constant length that depended on the bulk
concentration of motors.  This work demonstrated that motor-dependent
regulation of dynamics and length can occur not just for single MTs,
but for overlapping MT pairs.

Here we model the motor density profiles of motor proteins on
antiparallel MT overlaps of fixed lengths. We do not explicitly
consider MT length regulation in this system, as the combination of
binding kinetics and coupled switching of the motors between the two
antiparallel filaments is sufficient to produce a rich phenomenology.
Our work is an extension of Parmeggiani, Franosch, and Frey's TASEP
with binding kinetics on a single filament
\cite{parmeggiani_totally_2004,parmeggiani_phase_2003} to include two
antiparallel filaments coupled by switching; it is also an extension
of TASEP models of two antiparallel lanes with lane switching
\cite{juhasz_weakly_2007,ashwin_queueing_2010} to incorporate binding
and unbinding kinetics.

We first develop the model and show that measured and estimated
parameters can give overlap motor density profiles and motor
trajectories qualitatively similar to those found in the BTS
experiments. We then develop an analytic solution to the mean-field
steady state equations, and use it to determine the phase diagram of
the model. For high motor switching rate between filaments, we find a
new low density-high density-low density-high density phase. We then
study the model for the reference parameter set in more detail.
Because the motor density profiles are controlled by both boundary
conditions and a non-local total binding constraint, the density in
the center of the overlap depends not just on motor binding kinetics
but also on motor speed and overlap length.  We find an analytical
approximation that describes the overlap center density and the size
of the motor-dense boundary layer near the overlap ends. The degree of
motor accumulation near the overlap ends depends on the overlap length
and motor switching rate in addition to the motor speed and bulk
concentration.  The coupling of motor motion, binding, and switching
kinetics on antiparallel MT overlaps may therefore offer a previously
unrecognized mechanism for the control of one-dimensional motor
density profiles.

\section*{Materials and Methods}

\subsection*{Model}

\begin{figure}[t]
\begin{centering}
\includegraphics[width=0.95 \textwidth]{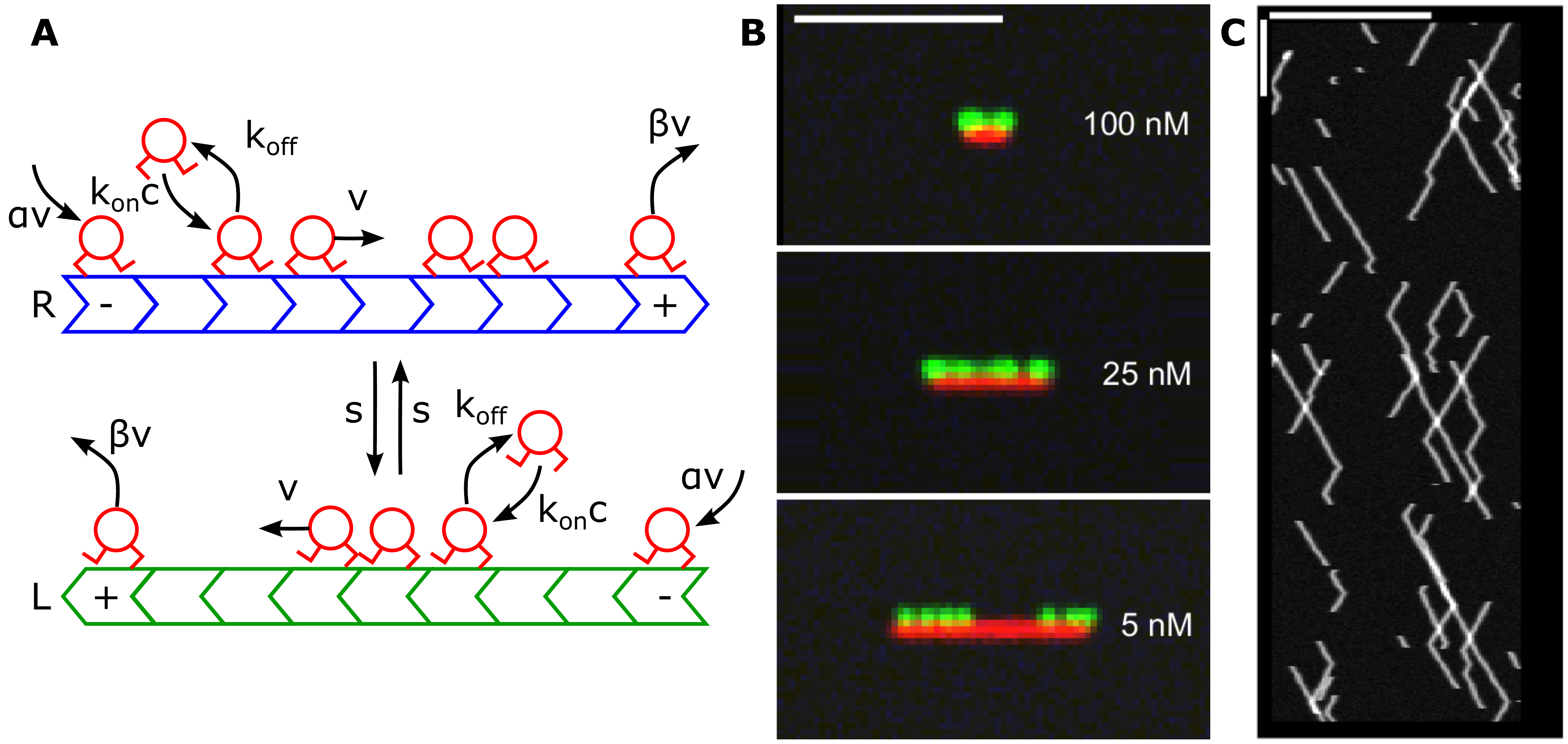}
\caption{Model and results overview. (A) Schematic of the model of
  motor motion on an antiparallel microtubule overlap. Two filaments
  (green and blue) are modeled as 1D lattices with their plus ends
  oppositely oriented. The filaments are labled R (L) if the plus end
  is pointing to the right (left). Motors (red) bind to empty lattice
  sites with rate $\konc$ and unbind with rate $\koff$. Bound motors
  step toward the MT plus end with rate $v$ (if the adjacent site
  toward the MT plus end is empty) or switch to the other MT with rate
  $s$ (if the corresponding site on the adjacent MT is empty). At MT
  minus ends, motors are inserted at rate $\alpha v$. At MT plus ends,
  motors are removed at rate $\beta v$. (B) Simulated experimental
  images made from our kMC model. Green, motor density. Red, overlap
  region. Scale bar, 5 $\mu$m. Simulations used to generate these
  images used the reference parameter set (table \ref{tab:units}) and
  the indicated bulk motor concentrations. (C) Simulated kymograph
  made from our kMC model with motor spatial position on the
  horizontal axis and time increasing down. Horizontal scale bar, 10
  $\mu$m. Vertical scale bar, 5s. The simulations used to generate the
  kymograph used the reference parameter set and 0.5 nM bulk motor
  concentration.}
\label{cartoon}
\end{centering}
\end{figure}

In this section, we develop a mathematical model of motor density on
antiparallel filament overlaps, inspired by the experiments of
Bieling, Telley, and Surrey (BTS) \cite{bieling_minimal_2010}. In the
BTS experiments, the crosslinking protein PRC1 binds preferentially to
overlapping regions of antiparallel MTs and recruits the kinesin motor
protein Xklp1 to the overlap region. Because PRC1 and therefore the
motors are present at much higher concentrations in the overlap, in
the model we consider the overlap region only and consider the regions
of single filaments as sources or sinks of motor proteins
(fig.~\ref{cartoon}A).  While previous work has shown that PRC1/Ase1
alone can develop density inhomogeneities and exert forces on sliding
MTs \cite{braun_adaptive_2011,lansky_diffusible_2015}, in the BTS
experiments little to no MT sliding occurred and the PRC1 distribution
was uniform in the overlaps \cite{bieling_minimal_2010}. Therefore we
do not explicitly model the PRC1 molecules or their spatial
distribution, but assume they are uniformly distributed so that each
lattice site in the overlap is identical with a binding affinity that
would correspond to the density-weighted average of the PRC1 and bare
tubulin affinities.  We treat each MT as a single track and neglect
the multiple-protofilament structure of the MT, consistent with
previous theoretical work
\cite{govindan_length_2008,varga_kinesin8_2009,hough_microtubule_2009,reese_crowding_2011,johann_length_2012,melbinger_microtubule_2012,kuan_biophysics_2013,reese_molecular_2014}.
In our model, motors can bind to and unbind from each of the MTs, walk
toward the plus end of a MT, and switch between the two MTs
(fig.~\ref{cartoon}A).

There are two competing processes in this model: motor stepping
(TASEP) and motor binding, unbinding, and MT switching (Langmuir
kinetics).  The Langmuir kinetics dominate the system behavior if the
overlap length is sufficiently large, $N\gg 1$, so that each bound
motor can only walk for a short fraction of the overlap length before
it unbinds. Thus, the competition between TASEP and Langmuir kinetics
occurs only if the overall binding rate to one MT in the overlap
$\Konc = N\konc$ (where $c$ is the bulk motor concentration), and
unbinding rate $\Koff = N\koff$, are of similar magnitude to the motor
speed $v$ \cite{parmeggiani_totally_2004}. 

In the next section, we develop the discrete microscopic model and
corresponding stochastic simulation, then derive the mean-field
continuum description. 

\subsubsection*{Discrete model}

We consider two overlapping antiparallel MTs of fixed length, so the
number of sites $N$ is fixed (fig.~\ref{cartoon}A). At each site,
motor binding or unbinding can occur with binding rate $\konc$, where
$\kon$ is the binding rate constant per site and $c$ the bulk motor
concentration, and unbinding rate $\koff$. Each bound motor steps at
rate $v$ to the next site toward the MT plus end (if the next site is
unoccupied), and switches at rate $s$ to the site on the adjacent MT
(if that site is unoccupied). Note that Xklp1 is plus-end directed,
and we therefore assume that the motion on a single MT is
unidirectional.

The occupation number $\hat{n}_i$ is $1$ if site $i$ is occupied or
$0$ if site $i$ is empty.  Then the equations for interior sites
($2<i<N-1$) on MTs with plus end to the right ($R$) and left ($L$) are
\begin{eqnarray}
  \nonumber
  \frac{d \hat{n}_{R, i}(t)}{dt} =&& v\hat{n}_{R, i-1}(t)[1-\hat{n}_{R,
                                     i}(t)]-v\hat{n}_{R,
                                     i}(t)[1-\hat{n}_{R, i+1}(t)] +\konc
                                     [1-\hat{n}_{R, i}(t)] \\ 
                                  &&-\koff \hat{n}_{R, i}(t) - s \hat{n}_{R,
                                     i}(t)[1-\hat{n}_{L, i}(t)] + s
                                     \hat{n}_{L, i}(t)[1-\hat{n}_{R, 
                                     i}(t)],\\ 
  \nonumber
  \frac{d \hat{n}_{L, i}(t)}{dt} =&& v\hat{n}_{L, i+1}(t)[1-\hat{n}_{L,
                                     i}(t)]-v\hat{n}_{L,
                                     i}(t)[1-\hat{n}_{L, i-1}(t)] +\konc
                                     [1-\hat{n}_{L, i}(t)] \\ 
                                  &&-\koff \hat{n}_{L, i}(t) - s \hat{n}_{L,
                                     i}(t)[1-\hat{n}_{R, i}(t)] +
                                     s\hat{n}_{R, i}(t)[1-\hat{n}_{L, 
                                     i}(t)].
\label{Fock_space_formalism}
\end{eqnarray}
At the boundary sites, we modify these equations to incorporate fluxes
into and out of the overlap.  The flux into the overlap is
$v \alpha[1-\hat{n}_1(t)]$, where nonzero $\alpha$ results from motors
moving into the overlap from the adjacent single-MT region. The flux
out of the overlap is $v \beta \hat{n}_N(t)$, where $\beta$ is derived
fromthe rate at which motors at an MT plus end unbind.  In principle,
each filament could have different boundary conditions. Because we
have no physical reason to distinguish the two halves of the overlap,
we focus on the symmetric case $\alpha_L=\alpha_R=\alpha$ and
$\beta_L=\beta_R=\beta$.  The resulting boundary site equations are
\begin{eqnarray}
  \frac{d \hat{n}_{R, 1}(t)}{dt}&=& v\alpha[1-\hat{n}_{R, 1}(t)]-
     v\hat{n}_{R, 1}(t)[1-\hat{n}_{R, 2}(t)],\\ 
  \frac{d \hat{n}_{R, N}(t)}{dt} &=& v\hat{n}_{R,
                                     N-1}(t)[1-\hat{n}_{R,
                                     N}(t)]-v\beta\hat{n}_{R, 
     N}(t),\\ 
  \frac{d \hat{n}_{L, 1}(t)}{dt} &=& v\hat{n}_{L, 2}(t)[1-\hat{n}_{L,
                                     1}(t)]-v\beta\hat{n}_{L, 
     1}(t), \\
  \frac{d \hat{n}_{L, N}(t)}{dt} &=& v\alpha[1-\hat{n}_{L, N}(t)]-
     v\hat{n}_{L, N}(t)[1-\hat{n}_{L, N-1}(t)].
\label{Fock_space_formalism_boundaries}
\end{eqnarray}
As shown below, these boundary conditions fix the motor densities to
be $\alpha$ at the minus end and $1-\beta$ at the plus end of each
filament.

At steady state, we can derive a total binding constraint on the
equations by summing all of sites on both filaments. The bulk flux
terms of the form $\hat{n}_{i-1}(t)[1-\hat{n}_{i}(t)]$ sum to zero and
only the binding and boundary terms remain:
\begin{equation}
  \sum_{i=2}^{N-1} [2\konc-(\konc+\koff)(\hat{n}_{R,i} +
  \hat{n}_{L,i})] +v\alpha[2-\hat{n}_{R, 1}(t)-\hat{n}_{L,
    N}(t)]-v\beta[\hat{n}_{R,N}(t)+\hat{n}_{L, 1}(t)]=0. 
\label{global_property}
\end{equation}
We find a binding constraint on the total motor binding
\begin{equation} 
  \sum_{i=1}^N \hat{n}_{R,i} +\hat{n}_{L,i}=2N
  \frac{\konc}{\konc+\koff} +
  \frac{2v[\alpha(1-\alpha)-\beta(1-\beta)]}{\konc+\koff}   = 2 N
  \rho_0 + \frac{2v[\alpha(1-\alpha)-\beta(1-\beta)]}{\konc+\koff},
  \label{eq:constraint}
\end{equation}
where we have defined the Langmuir density
$\rho_0=\konc/(\konc+\koff)$. Therefore, at steady state an
equilibrium involving binding, unbinding, and the filament-end
boundary conditions must be reached on average for the entire
overlap. This is related to the zero-current condition found in
previous work on the antiparallel TASEP without binding kinetics
\cite{juhasz_weakly_2007,ashwin_queueing_2010}.  We did not find an
analytical solution to the discrete equations.  Instead, we performed
kinetic Monte Carlo (kMC) simulations of the discrete model
(Supporting Material) and compared to solutions in the continuum
limit.

\subsubsection*{Mean-field continuum model}

We derive the mean-field continuum model as in previous work
\cite{parmeggiani_totally_2004} by taking the stationary average
$\langle\hat{n}_i\rangle \equiv \rho_i$, applying the random phase
approximation
$\langle\hat{n}_i\hat{n}_{i+1}\rangle =
\langle\hat{n}_i\rangle\langle\hat{n}_{i+1}\rangle$.
We also assume motor commutation during track switching
$\langle \hat{n}_{R,i} \hat{n}_{L,i}\rangle =\langle
\hat{n}_{L,i}\rangle \langle \hat{n}_{R,i}\rangle$
to give discrete mean-field equations with linear switching terms
(Supporting Material).  We take the continuum limit and
nondimensionalize the parameters and variables by choosing the length
of the overlap, $L$, as the unit of length and $L/v$ as the unit of
time.  Capital letters denote the nondimensionalized parameters
($S=sL/v$ and so on, Supporting Material).  We choose $x=0$ as the
center of the overlap, and the boundary conditions become
$\rho_R(-0.5) = \rho_L(0.5) = \alpha$ and
$\rho_R(0.5)=\rho_L(-0.5)=1-\beta$. The steady-state continuum
mean-field equations are
\begin{eqnarray}
\label{eq:cont1}
0&=& (2\rho_R-1) \frac{\partial \rho_R}{\partial x} + \Konc (1-\rho_R)
     - \Koff \rho_R -S \rho_R +S \rho_L , \\
\label{eq:cont2}
0&=&(1-2\rho_L) \frac{\partial \rho_L}{\partial x} + \Konc (1-\rho_L)
     - \Koff \rho_L +S \rho_R -S \rho_L. 
\end{eqnarray}
In the continuum mean-field model, the total binding constraint
(Eqn. \ref{eq:constraint}) is
\begin{equation}
  \label{eq:contin_constr}
  \int_{-1/2}^{1/2}dx\ \rho_R=\int_{-1/2}^{1/2}dx\ \rho_L =  \frac{
    \Kon c}{\Kon c + \Koff} + \frac{ \alpha(1-\alpha) -
    \beta(1-\beta)}{L(\Kon c + \Koff)}=\rho_0 + \frac{
    \alpha(1-\alpha) - \beta(1-\beta)}{L(\Kon c + \Koff)}. 
\end{equation}

\section*{Results}
\begin{table}
\footnotesize
  \begin{center}
    \begin{tabular}{|cp{4cm}p{3.5cm}p{6cm}|}
      \hline
      Symbol & Parameter & Reference value &  Notes  \\
      \hline
      $v$  & Motor speed & 0.5 $\mu$m\pers &   Measured by
                                             \citet{bieling_minimal_2010};
                                             varied up to 8 $\mu$m\pers\
                                             to study effect of
                                             varying speed on
                                             density profiles\\  
      $\kon$  & Binding rate constant   &
                                          $2.7\times10^{-4}$
                                          \pernm\pers &   Estimated
                                                        based on motor
                                                        density
                                                        profiles and
                                                        kymographs to 
                                                        give affinity
                                                        of 6.3 nM,
                                                        $\sim$15x
                                                        higher than
                                                        measured for
                                                        motor on
                                                        single MT by
                                                        \citet{bieling_microtubule_2010} \\ 
      $c$  & Bulk motor concentration & 1--200 nM &  Varied
                                                    by
                                                    \citet{bieling_minimal_2010}\\ 
      $\koff$  & Unbinding rate &0.169\pers &  Measured
                                              by
                                              \citet{bieling_minimal_2010} \\
      $s$  & Switching rate & $0.44$ \pers& Measured by \citet{bieling_minimal_2010} \\  
      $\alpha$  &Motor flux constant into overlap from MT minus end & 0 &
                                                                          Motors
                                                                          bind
                                                                          primarily
                                                                          inside
                                                                          the
                                                                          overlap;
                                                                          see
                                                                          discussion
                                                                          in
                                                                          the
                                                                          Results. We
                                                                          varied
                                                                          $\alpha$
                                                                          between
                                                                          0 and 1 to determine the model phase diagram\\  
      $\beta$  &Motor flux constant out of overlap from MT plus end & 0 &
                                                                          An
                                                                          upper
                                                                          bound
                                                                          on
                                                                          the
                                                                          end
                                                                          motor
                                                                          unbinding
                                                                          rate is $\beta =
                                                                          2.7 \times
                                                                          10^{-3}$;
                                                                          see
                                                                          discussion
                                                                          in
                                                                          the Results. We varied
                                                                          $\beta$
                                                                          between
                                                                          0 and 1 to determine the model phase diagram\\   
      $N$    & Number of sites & 120--2500& Varied to study overlaps of
                                            length 1--20 $\mu$m \\
      $\delta$    &  Length of a single site & 8 nm & Length of an
                                                      $\alpha$-$\beta$
                                                      tubulin
                                                      dimer, see \citet{bray_cell_2000}\\
      \hline
    \end{tabular}
    \caption{Parameter values for the reference parameter set, taken
      from experimental measurements or estimated as noted.}
    \label{tab:units}
  \end{center}
\end{table}

\subsection*{Motor density and trajectories in overlap}

To study whether our model can qualitatively describe the motor
density in antiparallel MT overlaps observed experimentally, we
determined a reference parameter set corresponding to the BTS
experiments (table \ref{tab:units}). Most parameters were directly
measured by BTS. We estimated the motor on rate constant $\kon$ by
comparison to the single-molecule imaging of low-density Xklp1 in an
overlap \cite{bieling_minimal_2010}. To estimate the minus-end
boundary condition $\alpha$, note that BTS found that Xklp1 binding
was greatly increased on overlaps due to recruitment by PRC1. In the
BTS experiments, low ionic strength (which favors motor-MT binding)
was needed to observe significant Xklp1 binding to single MTs outside
of overlaps. Therefore, we assume that the flux of motors into the
overlap from outside is negligible, and set $\alpha=0$ in our
reference parameters. The plus-end boundary condition $\beta$ is
related to the motor unbinding rate at MT plus ends. While this end
off rate was not directly measured by BTS, kinesin motors typically
pause at MT plus ends and have an end unbinding rate smaller than the
unbinding rate in the bulk. We therefore expect that $\beta$ lies
between 0 (no end unbinding) and $2.7 \times 10^{-3}$ (the value of
$\beta$ corresponding to an unbinding rate equal to $\koff$, the motor
unbinding rate in the bulk). We found that
$\beta = 2.7 \times 10^{-3}$ is so small that the motor density
profiles were indistinguishable from those with $\beta=0$
(fig.~\ref{fig:beta}). Therefore we used $\beta = 0$ in our reference
parameter set.

With these parameters, we studied kMC simulations of our model with
varying motor concentration and overlap lengths corresponding to the
steady-state values measured by BTS. We made simulated images that
represent how our model's motor distributions would appear in an
experiment with fluorescently tagged motors (green) and antiparallel
MT overlap region (red) (fig.~\ref{cartoon}B, Supporting
Material). Motors decorating the overlap, with greater density and end
accumulation for higher bulk motor concentration. The simulated images
are qualitatively similar to the experimental images (fig.~4 in
\citet{bieling_minimal_2010}). We also made simulated kymographs
(fig.~\ref{cartoon}C, Supporting Material), which show directed motor
motion with direction reversal, qualitatively similar to the
experimental results (figs.~6 and S5 in
\citet{bieling_minimal_2010}). We note that the experimental
kymographs show a larger population of paused/immobile motors than
occurs in our model.

\subsection*{Analytic solution of the steady-state continuum
  equations}

To further study the behavior of our model, we determined analytic
steady-state solutions of the continuum mean-field equations. One
solution to Eqn. \ref{eq:cont1}, \ref{eq:cont2} is the constant
Langmuir density set by binding/unbinding equilibrium,
$\rho_0=\Konc/(\Konc+\Koff)$. To find spatially varying solutions, we
first define $\sigma_{R,L}=\rho_{R,L} -\frac{1}{2} $, the difference
of the motor occupancies from $\frac{1}{2}$.  The rate combinations
are $k = \Konc+\Koff+S$ and $\gamma = \Konc-\Koff$. Then the equations
become
\begin{eqnarray}
\frac{d \sigma_R}{d x} &=& \frac{k}{2} - \frac{\gamma}{4
   \sigma_R}-\frac{S \sigma_L}{2 \sigma_R} 
\label{flow_q+}\\ 
\frac{d \sigma_L}{d x} &=& -\frac{k}{2} + \frac{\gamma}{4
   \sigma_L}+\frac{S \sigma_R}{2 \sigma_L}. 
\label{flow_q-}
\end{eqnarray}
Eqns \ref{flow_q+} and \ref{flow_q-} are well defined for
$\sigma_{R,L}\neq 0$.  This allows solution of the differential
equation relating $\sigma_R+\sigma_L$ and $\sigma_R-\sigma_L$ with one
integration constant (Supporting Material).  Our kMC simulation
results for motor density in the overlap differ substantially from
this analytic solution (fig.~\ref{fig:phase}), because the analytic
solution (Eqn.~\ref{exact}) does not satisfy the total binding
constraint (Eqn.~\ref{eq:contin_constr}).  Therefore, the full density
profiles must be determined by matching continuum solutions
corresponding to different integration constants. This mathematical
result implies several remarkable properties of the density profiles,
as discussed below.

\subsection*{Phase diagram}

\begin{figure}[t]
\begin{centering}
\includegraphics[width=0.5
\textwidth]{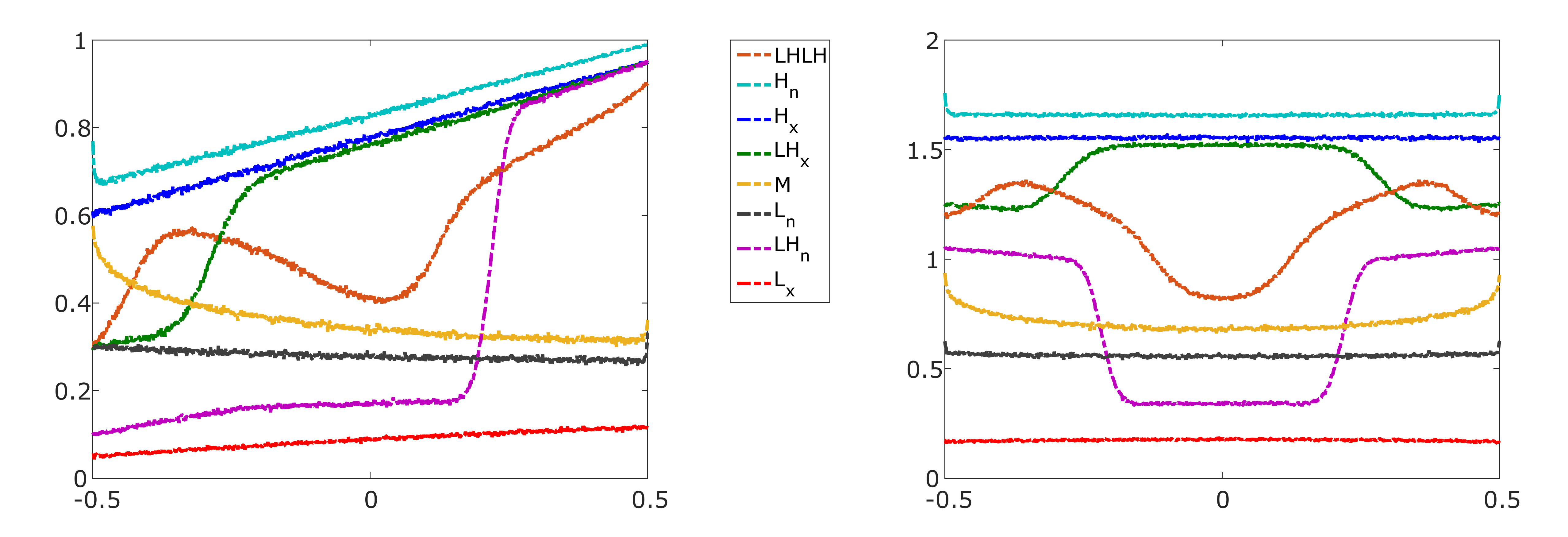}\includegraphics[width=0.4
\textwidth]{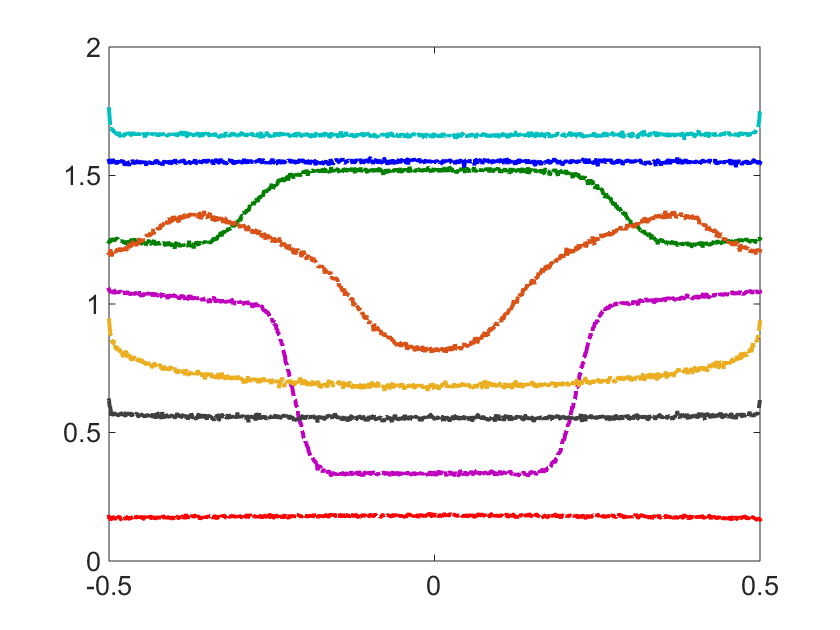}
\caption{Nonequilibrium phases.  Left: motor density $\rho_R (x)$ on
  MT with rightward-moving motors. Right: total motor density
  $\rho_R(x) + \rho_L (x)$ on both MTs in the overlap.  The parameters
  used are the reference parameter set of table \ref{tab:units} with a
  bulk motor concentration of 200 nM, except for the switching rate
  which is 0.5 \pers\ for the LHLH curve and 0.1 \pers\ for all other
  curves and the boundary conditions as noted below. LHLH: low
  density-high density-low density-high density coexistence
  ($\alpha=0.3$, $\beta=0.9$, orange). H$_{\rm e}$: high-density
  end-accumulating phase ($\alpha=0.95$, $\beta=0.99$, light blue).
  H$_{\rm c}$: high-density center-accumulating phase ($\alpha=0.6$,
  $\beta= 0.95$, dark blue). LH$_{\rm c}$: low density-high density
  center-accumulating coexistence ($\alpha=0.3$, $\beta=0.95$, green).
  M: Meissner phase $\alpha=0.6$, $\beta=0.4$, yellow).  L$_{\rm e}$:
  low-density end-accumulating phase ($\alpha=0.3$, $\beta=0.4$,
  black).  LH$_{\rm e}$: low density-high density end-accumulating
  coexistence ($\alpha=0.1$, $\beta= 0.95$, magenta).  L$_{\rm c}$:
  low-density center-accumulating phase ($\alpha=0.05$, $\beta= 0.1$,
  red).  }
\label{fig:phases}
\end{centering}
\end{figure}

We determined the phase diagram of the model as a function of the
boundary condition parameters $\alpha$ and $\beta$, as well as how the
phase diagram changes with the other model parameters.  We find four
phases previously observed for the single-filament case
\cite{parmeggiani_totally_2004}, the low density (L), high density
(H), low density-high density (LH), and Meissner (M) phases
(figs.~\ref{fig:phases}, \ref{nonlinear_flow_phases}). We distinguish
two different classes of L, H, and LH phase on overlaps: the
center-accumulating case (abbreviated c) has a higher total motor
concentration profile near the center of the filament and motor
accumulation at the overlap center, while the end-accumulating case
(e) has a lower total motor concentration profile near the filament
center and motor accumulation at the overlap ends. In addition, for
high switching rate we observe a new low density-high density-low
density-high density (LHLH) phase (fig.~\ref{fig:phases}).

\begin{figure}[t]
\begin{centering}
\includegraphics[width=0.45 \textwidth]{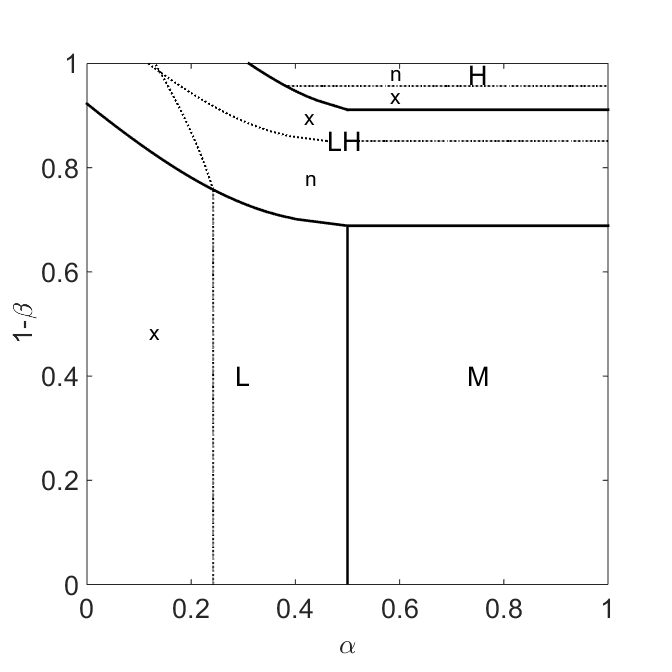}
\includegraphics[width=0.45 \textwidth]{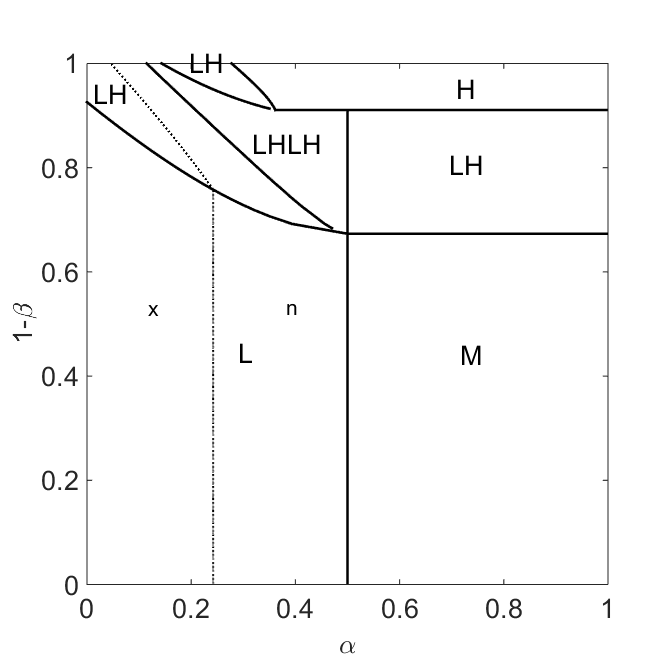}
\caption{Phase diagrams. Left: low switching rate (0.1 \pers); right:
  high switching rate (0.5 \pers). Solid lines indicate phase
  boundaries, and dotted lines boundaries between center-accumulating
  (v) and end-accumulating (e) overlap motor density profiles.  The
  bulk motor concentration is 200 nM, the motor speed is 5 \mum\pers,
  and other parameters are the reference values (table
  \ref{tab:units}).}
\label{phase_diagram}
\end{centering}
\end{figure}

We studied flows in the $\sigma_L-\sigma_R$ phase plane to determine
the motor concentration profiles
(fig.~\ref{nonlinear_phase_space_flow}) and therefore the phase
diagram, as discussed in the Supporting Material.  Representative
phase diagrams for small and large switching rate are shown in
fig.~\ref{phase_diagram}.  We note that the new LHLH phase occurs when
$\alpha < 0.5$ and the switching frequency is sufficiently high,
conditions which apply to the BTS experiments. This phase can appear
for high switching rate when two analytic solutions to the
steady-state continuum equations intersect, causing transition points
in the phase plane where the density is ill defined
(fig.~\ref{nonlinear_phase_space_flow}--\ref{exact_solution_beyond_and_above_transition_lin}). Our
LHLH phase is reminiscent of multi-phase coexistence found by Pierobon
\cite{pierobon_bottleneckinduced_2006}. However, in this previous work
the multi-phase co-existence occurs due to a point defect in the lane,
while here it occurs for spatially uniform dynamics.

The phase boundaries change with parameters of the model. Increases
(decreases) to the Langmuir density (through changes in bulk motor
concentration or binding/unbinding rate constants) shift the lower
boundary of the LH phase closer to (farther from) the line where
$1-\beta = 0.5$ when $\alpha > \rho_0$ and closer to (farther from)
the line $\alpha = \beta$ when $\alpha < \rho_0$. Similar shifts occur
for the upper boundary of the LH phase. The motor speed affects the
width and the shape of the LH phase: for higher motor speed, the LH
phase region becomes narrower in width and the LH boundaries become
more flat.

\subsection*{Density profiles for reference parameters}

\begin{figure}[t]
\begin{centering}
\includegraphics[width=0.45 \textwidth]{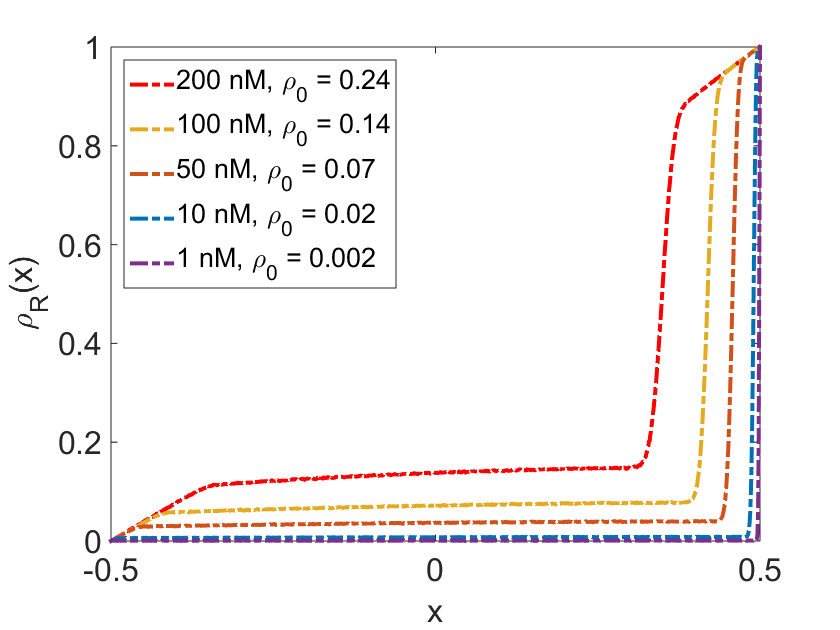}
\includegraphics[width=0.45 \textwidth]{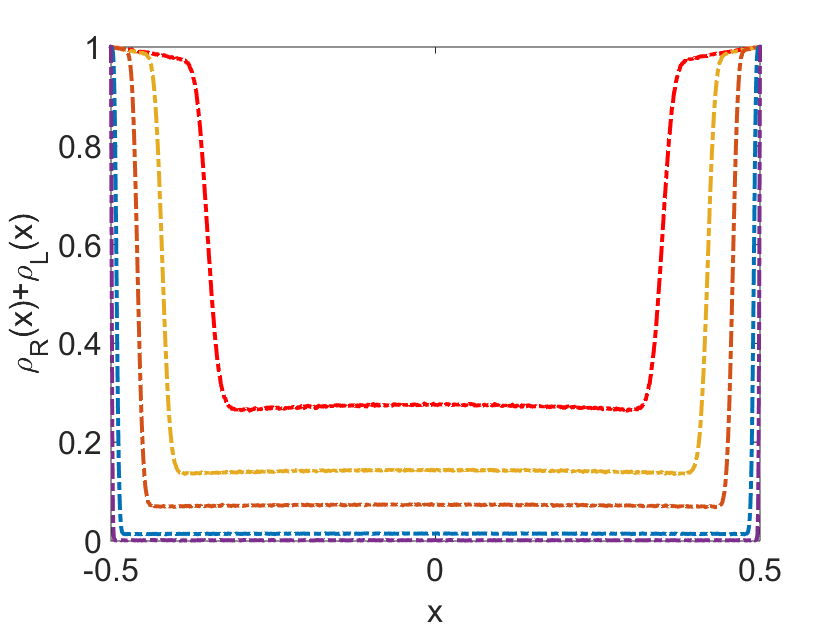}
\caption{Motor density profiles for the reference parameter set. Left:
  motor density $\rho_R (x)$ on MT with rightward-moving
  motors. Right: total motor density $\rho_R(x) + \rho_L (x)$ on both
  MTs in the overlap. Parameters are the reference parameter set of
  table \ref{tab:units} with the bulk motor concentrations indicated
  in the legend. }
\label{fig:density_x}
\end{centering} 
\end{figure}

The motor density profiles that we determine by kMC simulation are
qualitatively consistent with those observed experimentally
(figs.~\ref{cartoon}, \ref{fig:density_x}). Here we illustrate our
results for the reference parameter set (table \ref{tab:units}); in
the Supporting Material we also discuss a large-$S$ parameter set
chosen to approximate the limit of large switching rate (table
\ref{tab:units_extended}, fig.~\ref{fig:density_x_larges})). This
parameter regime typically shows the LH phase, where the density
profiles have three regions separated by domain walls.  For motors
moving to the right, there is a boundary layer on the left with the
motor density increasing from zero, then a region of approximately
linearly varying density, then a sharp transition to another boundary
layer of linearly increasing density that approaches 1 on the right
boundary (fig.~\ref{fig:density_x}).  If we fix all other parameters,
the boundary layer regions increase in size as the bulk motor
concentration increases. The total motor density in the overlap is the
sum of the motor densities on the two filaments, which have the
symmetry $\rho_L(-x)=\rho_R(x)$.  The linear density variation near
the overlap ends can be seen in the continuum mean-field equations
(Supporting Material): the slope is $\Kon c + S$ at the minus end and
$\Koff + S$ at the plus end. This approximation agrees well with
simulation results near the overlap ends (fig.~\ref{fig:phase},
\ref{linsoln}).

\subsection*{Control of center density by the total binding
  constraint}

Motor density profiles must satisfy the total binding constraint of
Eqn.~\ref{eq:contin_constr}, which requires that when
$\alpha = \beta = 0$, the \textit{integral} of the density on a single
MT must equal $\rho_0$, the Langmuir density determined by
binding/unbinding equilibrium. We verified that the total binding
constraint is satisfied in our simulations by determining the
integrated motor density and comparing it to $\rho_0$
(fig.~\ref{overall_conc}).  However, as noted above, the analytic
solutions we found for the steady-state motor density
(Eqns.~\ref{exact}, \ref{nearend1}, \ref{nearend2}) do not in general
satisfy the total binding constraint.  As a result, the motor
concentration in the center of the filaments does not necessarily
approach the Langmuir density $\rho_0=\Kon c/(\Kon c + \Koff)$
(fig.~\ref{fig:density_x}). This is a significant difference between
the model of microtubule overlaps we study and the single-filament
model \cite{parmeggiani_totally_2004}.

The total binding constraint also controls the length of the boundary
layer regions near the overlap ends.  To derive an analytic
approximation expression for this length, we approximate the motor
densities as piecewise linear, and assume the domain walls are
infinitely thin so that we can neglect them in integrating the
concentration (fig.~\ref{density_regions}). A filament is divided into
three regions: boundary layers near the plus and minus ends, and the
central region. We can then determine the motor density profile
(Eqn.~\ref{eq:rho_lin2}), the boundary layer ends, $x_{bl}$
(Eqn.~\ref{eq:xbl}), and the motor density in the center of the
overlap, $\rho_c$ (Eqn.~\ref{eq:center_dens}).

\begin{figure}[t]
\begin{centering}
\includegraphics[width=0.45 \textwidth]{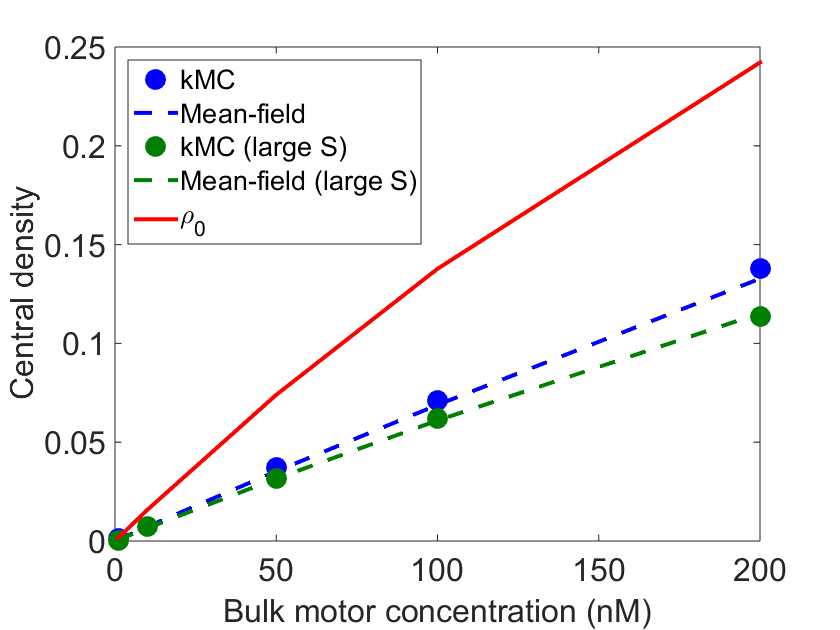}
\includegraphics[width=0.45 \textwidth]{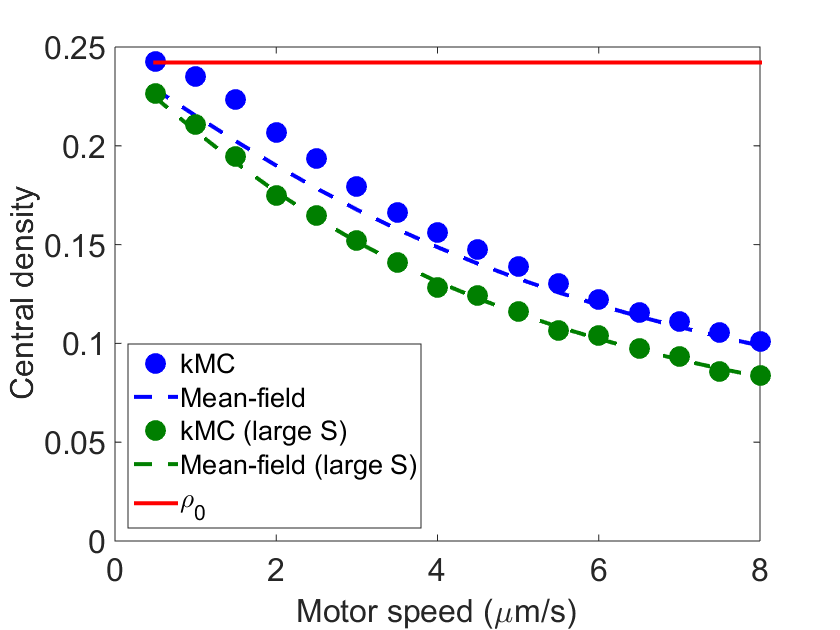}
\caption{Motor density at the center of the overlap. Left: variation
  with bulk motor concentration; right: variation with motor
  speed. Points indicate simulation results and dashed lines
  theoretical prediction from Eqn.~\ref{eq:center_dens}. The red line
  shows the value of the Langmuir density $\rho_0$ for the same
  parameters.}
\label{fig:center_conc}
\end{centering}
\end{figure}
\begin{figure}[t]
\begin{centering}
\includegraphics[width=0.45 \textwidth]{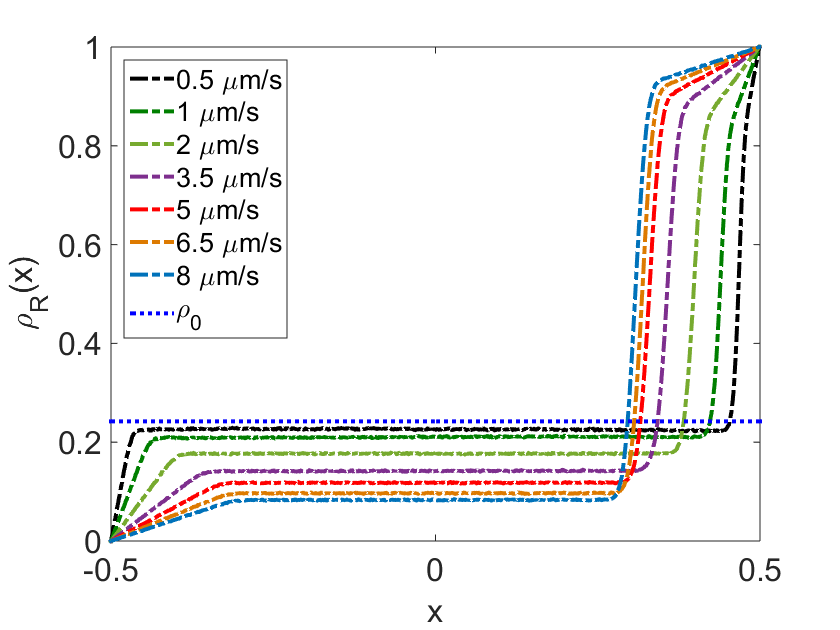}
\includegraphics[width=0.45 \textwidth]{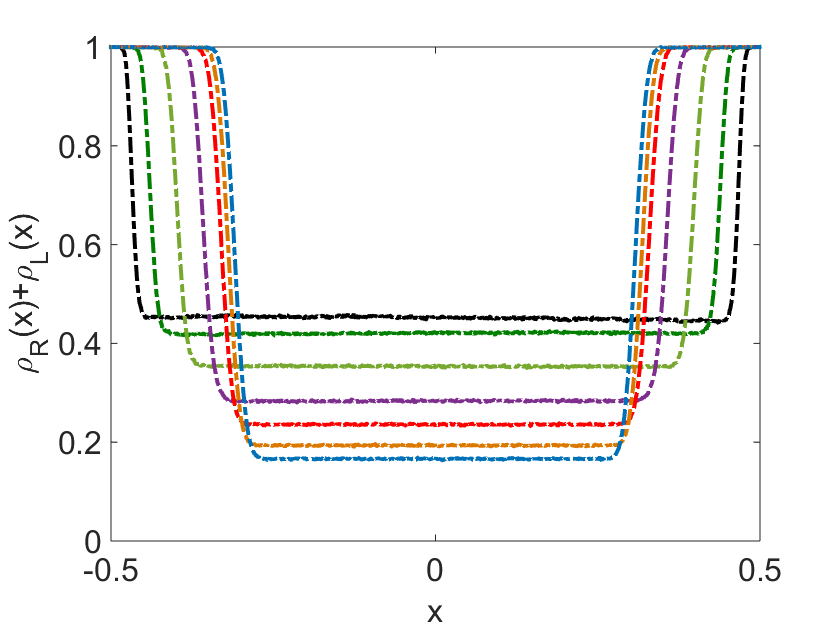}
\caption{Dependence of density profiles on motor speed. Left: motor
  density $\rho_R (x)$ on MT with rightward-moving motors. Right:
  total motor density $\rho_R(x) + \rho_L (x)$ on both MTs in the
  overlap.  Varying motor speed can significantly alter the motor
  density at the center of the overlap.  These simulations use the
  large-$S$ parameter set of table \ref{tab:units_extended} with bulk
  motor concentration $c=200$ nM and the motor speeds indicated in the
  legend.}
\label{fig:density_speed}
\end{centering}
\end{figure}

The motor density at the center of the overlap $\rho_c \neq \rho_0$,
even when the overlap is long. Instead the center density depends on
the motor speed and filament switching rate in addition to binding
parameters (fig.~\ref{fig:density_x}) In fig.~\ref{fig:center_conc},
we compare simulation results and the prediction of
Eqn.~\ref{eq:center_dens}, showing that our approximation is in good
agreement with simulations.  (Note that our approximation of the
density as linearly varying becomes less exact for slow motor speeds,
leading to a discrepancy between simulation results and our analytic
approximation.)  For comparison, we show the Langmuir density $\rho_0$
for the same parameter. Varying motor speed has no effect on $\rho_0$
(right panel), but a significant effect on $\rho_c$. To further
illustrate the dependence of the center density on motor kinetic
parameters, we show in fig.~\ref{fig:density_speed} examples of motor
density profiles from kMC simulations with varying motor speed.
Changing motor kinetic parameters can have a large effect on the
center density in the overlap, even for systems with
\textit{identical} binding kinetic parameters and $\rho_0$.

\subsection*{Control of end accumulation by the total binding
  constraint}

The length of the boundary layer at the overlap ends depends on motor
binding and kinetic parameters as well as the overlap length
(Eqn.~\ref{eq:xbl}). We compared the analytic approximation to
simulation results and found good agreement (fig.~\ref{bl_length}),
allowing us
to predict how varying experimental control parameters will alter the
degree of motor accumulation near the overlap ends
(fig.~\ref{predictions}). We note that although the boundary layer
lengths predicted for typical experimental parameters are below the
optical resolution limit of $\approx 250$ nm, they could be
distinguished by fits of fluorescent motor intensity profiles to model
predictions (c.f. figs.~\ref{cartoon}, \ref{fig:density_x},
\ref{fig:density_speed}), similar to how fits to a Gaussian
fluorescence intensity profile allow sub-resolution localization of
single molecules \cite{yildiz_myosin_2003}.

The total binding constraint means the the boundary layer length is
determined as fraction of the overlap length. In the single-filament
model, the motor density profile in the MT minus-end boundary layer is
independent of MT length
\cite{hough_microtubule_2009,kuan_biophysics_2013}, with important
consequences: MT length regulation by the kinesin-8 motor Kip3
\cite{varga_yeast_2006,varga_kinesin8_2009} depends on a motor density
profile that maintains the same functional form near the minus end as
MT length changes.  By contrast, on antiparallel MT overlaps the
density profile, boundary-layer length, and extent of motor
accumulation at the overlap ends change with overlap length
(fig.~\ref{predictions}). This means that a length regulation
mechanism identical to that of Kip3 cannot occur on antiparallel MT
overlaps.

\section*{Conclusion}

We studied a model of motor motion on antiparallel MT overlaps that
incorporates motor binding and unbinding, plus-end directed motor
motion, and switching between filaments (fig.~\ref{cartoon}). Our
model is inspired by the experiments of Bieling, Telley, and Surrey on
the motion of the kinesin-4 motor Xklp1 on antiparallel MT overlaps
\cite{bieling_minimal_2010}.  Our model is an extension of previous
theory that studied motor motion on a single MT with binding kinetics
\cite{parmeggiani_totally_2004,parmeggiani_phase_2003}, or motor
motion on two antiparallel filaments with lane switching, but no
binding kinetics \cite{juhasz_weakly_2007,ashwin_queueing_2010}.  To
our knowledge, this is the first theoretical study of a two-lane TASEP
model with oppositely oriented lanes, switching, and
binding/unbinding.

To compare to the BTS experiments, we used measured or estimated
parameters (table \ref{tab:units}). Because PRC1 recruits motors
directly into the overlap and motor binding to single MTs is much
weaker, we neglect motor binding to MTs outside of the overlap and set
the minus-end flux parameter $\alpha = 0$.  Using the bulk motor
unbinding rate $\koff$ as an upper bound on the plus-end unbinding
rate, we find $\beta \le 2.7 \times 10^{-3}$. This value is
sufficiently small that it gives motor density profiles
indistinguishable from those with $\beta=0$
(fig.~\ref{fig:beta}). Therefore no-flux boundary conditions with
$\alpha = \beta = 0$ are used to model the experiments.  Simulated
images of the motor distribution in an overlap and simulated
kymographs of motor trajectories are similar to those found
experimentally (fig.~\ref{cartoon}).

We derived analytical and approximate solutions of the continuum
steady-state equations and compared them to kMC simulation results
(fig.~\ref{fig:phase}). Focusing on the symmetric case where the
boundary conditions $\alpha$ and $\beta$ are the same for both
filaments in the overlap, we used both kMC simulations and phase plane
flows (fig.~\ref{nonlinear_flow_phases} --
\ref{exact_solution_beyond_and_above_transition_lin}) to determine the
nonequilibrium phases possible in our model
(fig.~\ref{fig:phases}). For low rate of motor switching between
filaments in the overlap, we find the low density, high density, low
density-high density, and Meissner phases previously studied for the
single-lane case \cite{parmeggiani_totally_2004}. In addition, for
high switching rate we find a novel low density-high density-low
density-high density phase with three domain walls. We determined
representative phase diagrams for small and large switching rate
(fig.~\ref{phase_diagram}).

We then studied the model in more detail both for the reference
parameters and a high-switching-rate parameter set (table
\ref{tab:units_extended}). Results of kMC simulations
(fig.~\ref{fig:density_x}, \ref{fig:density_x_larges}) agree well with
exact and approximate solutions of the continuum steady-state
equations (fig.~\ref{fig:phase}, \ref{linsoln}).  In contrast to
systems in which motors move on single filaments
\cite{govindan_length_2008,varga_kinesin8_2009,hough_microtubule_2009,reese_crowding_2011,johann_length_2012,melbinger_microtubule_2012,kuan_biophysics_2013,reese_molecular_2014},
we find that antiparallel overlaps with no-flux boundary conditions
have total zero current. This leads to a total binding constraint that
the integral of the total motor density on a single filament must
equal $\rho_0$, the motor density set by binding/unbinding equilibrium
(fig.~\ref{overall_conc}).

For the experimentally relevant low density-high density coexistence
phase, the density profiles are approximately piecewise linear
(fig.~\ref{density_regions}). This motivates an analytic approximation
to determine the density profiles consistent with the total binding
constraint and gives analytic expressions for the overlap center motor
density and the length of the boundary layer in which motors
accumulate near the overlap ends.  We find that as a result of the
total binding constraint, the motor density at the center of the
overlap is not determined solely by the motor binding equilibrium, but
is also controlled by the overlap length, motor speed, and filament
switching rate (fig.~\ref{fig:center_conc}, \ref{fig:density_speed}).
These same parameters control the length of the boundary layer at the
overlap ends where motors accumulate (fig.~\ref{bl_length}).

The mitotic spindle contains arrays of overlapping antiparallel MTs to
which multiple motors and crosslinkers bind
\cite{fededa_molecular_2012}. The surprising differences in motor
density profiles between single filaments and the antiparallel
overlaps we study here are therefore of interest in the study of the
spindle midzone.  For antiparallel overlaps, both the motor density
far from the overlap ends and the number of motors near the overlap
ends can be tuned not just by motor binding kinetics but also by motor
speed, filament switching rate, and overlap length
(fig.~\ref{predictions}). The antiparallel filament geometry gives
biological systems additional handles to control motor density, its
spatial distribution, and therefore motor function.  Motor density can
affect recruitment of other proteins and MT dynamics. Therefore, this
previously undescribed mechanism of regulation of motor density along
MTs may offer advantages to the control of motor activity.

For single MTs, length regulation by the kinesin-8 motor Kip3 depends
on a functional form of motor density that is independent of MT length
\cite{varga_yeast_2006,varga_kinesin8_2009}. The physics we describe
here means that the motor density and extent of motor accumulation at
the overlap ends depends on overlap length. Therefore, length
regulation of antiparallel MT overlaps must occur differently.  In
future work, it will be of interest to understand how the motor
density profiles on fixed-length overlaps can be used to understand
the length regulation of dynamic overlaps.

\section*{Acknowledgements}

We thank Robert Blackwell, Matthew Glaser, and Loren Hough for useful
discussions. This work was supported by NSF grant DMR-0847685 and NIH
grant GM110486 to MDB, fellowship to H-SK  provided by
matching funds from the NIH/CU Biophysics Training Program, and
facilities of the Soft Materials Research Center under NSF MRSEC
Grants DMR-0820579 and DMR-1420736.

\bibliography{references}{}

\begin{thebibliography}{41}
\providecommand{\url}[1]{\texttt{#1}}
\providecommand{\urlprefix}{ }

\bibitem[{Bray}(2000)]{bray_cell_2000}
{Bray}, D., 2000.
\newblock Cell movements: from molecules to motility.
\newblock {Routledge}.

\bibitem[{Kolomeisky}(2015)]{kolomeisky_motor_2015}
{Kolomeisky}, A.~B., 2015.
\newblock Motor {{Proteins}} and {{Molecular Motors}}.
\newblock {CRC Press}.

\bibitem[{Chowdhury}(2013)]{chowdhury_modeling_2013}
{Chowdhury}, D., 2013.
\newblock Modeling {{Stochastic Kinetics}} of {{Molecular Machines}} at
  {{Multiple Levels}}: {{From Molecules}} to {{Modules}}.
\newblock \emph{Biophysical Journal} 104:2331--2341.

\bibitem[{Helbing}(2001)]{helbing_traffic_2001}
{Helbing}, D., 2001.
\newblock Traffic and related self-driven many-particle systems.
\newblock \emph{Rev. Mod. Phys.} 73:1067--1141.

\bibitem[{Parmeggiani} et~al.(2004){Parmeggiani}, {Franosch}, and
  {Frey}]{parmeggiani_totally_2004}
{Parmeggiani}, A., T.~{Franosch}, and E.~{Frey}, 2004.
\newblock Totally asymmetric simple exclusion process with {{Langmuir}}
  kinetics.
\newblock \emph{Phys. Rev. E} 70:046101.

\bibitem[{Leduc} et~al.(2012){Leduc}, {Padberg-Gehle}, {Varga}, {Helbing},
  {Diez}, and {Howard}]{leduc_molecular_2012}
{Leduc}, C., K.~{Padberg-Gehle}, V.~{Varga}, D.~{Helbing}, S.~{Diez}, and
  J.~{Howard}, 2012.
\newblock Molecular crowding creates traffic jams of kinesin motors on
  microtubules.
\newblock \emph{PNAS} 109:6100--6105.

\bibitem[{Varga} et~al.(2009){Varga}, {Leduc}, {Bormuth}, {Diez}, and
  {Howard}]{varga_kinesin8_2009}
{Varga}, V., C.~{Leduc}, V.~{Bormuth}, S.~{Diez}, and J.~{Howard}, 2009.
\newblock Kinesin-8 {{Motors Act Cooperatively}} to {{Mediate Length-Dependent
  Microtubule Depolymerization}}.
\newblock \emph{Cell} 138:1174--1183.

\bibitem[{Bieling} et~al.(2010{\natexlab{a}}){Bieling}, {Telley}, and
  {Surrey}]{bieling_minimal_2010}
{Bieling}, P., I.~A. {Telley}, and T.~{Surrey}, 2010.
\newblock A {{Minimal Midzone Protein Module Controls Formation}} and
  {{Length}} of {{Antiparallel Microtubule Overlaps}}.
\newblock \emph{Cell} 142:420--432.

\bibitem[{Goshima} et~al.(2005){Goshima}, {Wollman}, {Stuurman}, {Scholey}, and
  {Vale}]{goshima_length_2005}
{Goshima}, G., R.~{Wollman}, N.~{Stuurman}, J.~M. {Scholey}, and R.~D. {Vale},
  2005.
\newblock Length control of the metaphase spindle.
\newblock \emph{Curr. Biol.} 15:1979--1988.

\bibitem[{Walczak} et~al.(1996){Walczak}, {Mitchison}, and
  {Desai}]{walczak_xkcm1_1996}
{Walczak}, C.~E., T.~J. {Mitchison}, and A.~{Desai}, 1996.
\newblock {{XKCM1}}: {{A Xenopus Kinesin-Related Protein That Regulates
  Microtubule Dynamics}} during {{Mitotic Spindle Assembly}}.
\newblock \emph{Cell} 84:37--47.

\bibitem[{Rivero} et~al.(1996){Rivero}, {Koppel}, {Peracino}, {Bozzaro},
  {Siegert}, {Weijer}, {Schleicher}, {Albrecht}, and
  {Noegel}]{rivero_role_1996}
{Rivero}, F., B.~{Koppel}, B.~{Peracino}, S.~{Bozzaro}, F.~{Siegert}, C.~J.
  {Weijer}, M.~{Schleicher}, R.~{Albrecht}, and A.~A. {Noegel}, 1996.
\newblock The role of the cortical cytoskeleton: {{F}}-actin crosslinking
  proteins protect against osmotic stress, ensure cell size, cell shape and
  motility, and contribute to phagocytosis and development.
\newblock \emph{J Cell Sci} 109:2679--2691.

\bibitem[{Revenu} et~al.(2004){Revenu}, {Athman}, {Robine}, and
  {Louvard}]{revenu_coworkers_2004}
{Revenu}, C., R.~{Athman}, S.~{Robine}, and D.~{Louvard}, 2004.
\newblock The co-workers of actin filaments: from cell structures to signals.
\newblock \emph{Nat. Rev. Mol. Cell Biol.} 5:635--646.

\bibitem[{Dogterom} and {Leibler}(1993)]{dogterom_physical_1993}
{Dogterom}, M., and S.~{Leibler}, 1993.
\newblock Physical aspects of the growth and regulation of microtubule
  structures.
\newblock \emph{Phys. Rev. Lett.} 70:1347--1350.

\bibitem[{Drummond}(2011)]{drummond_regulation_2011}
{Drummond}, D.~R., 2011.
\newblock Regulation of microtubule dynamics by kinesins.
\newblock \emph{Seminars in Cell \& Developmental Biology} 22:927--934.

\bibitem[{Gupta} et~al.(2006){Gupta}, {Carvalho}, {Roof}, and
  {Pellman}]{gupta_endspecific_2006}
{Gupta}, M.~L., P.~{Carvalho}, D.~M. {Roof}, and D.~{Pellman}, 2006.
\newblock Plus end-specific depolymerase activity of {{Kip3}}, a kinesin-8
  protein, explains its role in positioning the yeast mitotic spindle.
\newblock \emph{Nat Cell Biol} 8:913--923.

\bibitem[{Varga} et~al.(2006){Varga}, {Helenius}, {Tanaka}, {Hyman}, {Tanaka},
  and {Howard}]{varga_yeast_2006}
{Varga}, V., J.~{Helenius}, K.~{Tanaka}, A.~A. {Hyman}, T.~U. {Tanaka}, and
  J.~{Howard}, 2006.
\newblock Yeast kinesin-8 depolymerizes microtubules in a length-dependent
  manner.
\newblock \emph{Nat Cell Biol} 8:957--962.

\bibitem[{Hough} et~al.(2009){Hough}, {Schwabe}, {Glaser}, {McIntosh}, and
  {Betterton}]{hough_microtubule_2009}
{Hough}, L.~E., A.~{Schwabe}, M.~A. {Glaser}, J.~R. {McIntosh}, and M.~D.
  {Betterton}, 2009.
\newblock Microtubule depolymerization by the kinesin-8 motor {{Kip3p}}: a
  mathematical model.
\newblock \emph{Biophys. J.} 96:3050--3064.

\bibitem[{Reese} et~al.(2011){Reese}, {Melbinger}, and
  {Frey}]{reese_crowding_2011}
{Reese}, L., A.~{Melbinger}, and E.~{Frey}, 2011.
\newblock Crowding of {{Molecular Motors Determines Microtubule
  Depolymerization}}.
\newblock \emph{Biophys. J.} 101:2190--2200.

\bibitem[{Govindan} et~al.(2008){Govindan}, {Gopalakrishnan}, and
  {Chowdhury}]{govindan_length_2008}
{Govindan}, B.~S., M.~{Gopalakrishnan}, and D.~{Chowdhury}, 2008.
\newblock Length control of microtubules by depolymerizing motor proteins.
\newblock \emph{Europhys. Lett.} 83:40006.

\bibitem[{Johann} et~al.(2012){Johann}, {Erlenk{\"a}mper}, and
  {Kruse}]{johann_length_2012}
{Johann}, D., C.~{Erlenk{\"a}mper}, and K.~{Kruse}, 2012.
\newblock Length {{Regulation}} of {{Active Biopolymers}} by {{Molecular
  Motors}}.
\newblock \emph{Phys. Rev. Lett.} 108:258103.

\bibitem[{Melbinger} et~al.(2012){Melbinger}, {Reese}, and
  {Frey}]{melbinger_microtubule_2012}
{Melbinger}, A., L.~{Reese}, and E.~{Frey}, 2012.
\newblock Microtubule {{Length Regulation}} by {{Molecular Motors}}.
\newblock \emph{Phys. Rev. Lett.} 108:258104.

\bibitem[{Reese} et~al.(2014){Reese}, {Melbinger}, and
  {Frey}]{reese_molecular_2014}
{Reese}, L., A.~{Melbinger}, and E.~{Frey}, 2014.
\newblock Molecular mechanisms for microtubule length regulation by kinesin-8
  and {{XMAP215}} proteins.
\newblock \emph{Interface Focus} 4:20140031.

\bibitem[{Tischer} et~al.(2010){Tischer}, {ten Wolde}, and
  {Dogterom}]{tischer_providing_2010}
{Tischer}, C., P.~R. {ten Wolde}, and M.~{Dogterom}, 2010.
\newblock Providing {{Positional Information}} with {{Active Transport}} on
  {{Dynamic Microtubules}}.
\newblock \emph{Biophys. J.} 99:726--735.

\bibitem[{Kuan} and {Betterton}(2013)]{kuan_biophysics_2013}
{Kuan}, H.-S., and M.~D. {Betterton}, 2013.
\newblock Biophysics of filament length regulation by molecular motors.
\newblock \emph{Phys. Biol.} 10:036004.

\bibitem[{Glun{\v c}i{\'c}} et~al.(2015){Glun{\v c}i{\'c}}, {Maghelli},
  {Krull}, {Krsti{\'c}}, {Ramunno-Johnson}, {Pavin}, and
  {Toli{\'c}}]{gluncic_kinesin8_2015}
{Glun{\v c}i{\'c}}, M., N.~{Maghelli}, A.~{Krull}, V.~{Krsti{\'c}},
  D.~{Ramunno-Johnson}, N.~{Pavin}, and I.~M. {Toli{\'c}}, 2015.
\newblock Kinesin-8 {{Motors Improve Nuclear Centering}} by {{Promoting
  Microtubule Catastrophe}}.
\newblock \emph{Phys. Rev. Lett.} 114:078103.

\bibitem[{Kurasawa} et~al.(2004){Kurasawa}, {Earnshaw}, {Mochizuki}, {Dohmae},
  and {Todokoro}]{kurasawa_essential_2004}
{Kurasawa}, Y., W.~C. {Earnshaw}, Y.~{Mochizuki}, N.~{Dohmae}, and
  K.~{Todokoro}, 2004.
\newblock Essential roles of {{KIF4}} and its binding partner {{PRC1}} in
  organized central spindle midzone formation.
\newblock \emph{EMBO J.} 23:3237--3248.

\bibitem[{Zhu} and {Jiang}(2005)]{zhu_cell_2005}
{Zhu}, C., and W.~{Jiang}, 2005.
\newblock Cell cycle-dependent translocation of {{PRC1}} on the spindle by
  {{Kif4}} is essential for midzone formation and cytokinesis.
\newblock \emph{PNAS} 102:343--348.

\bibitem[{Khmelinskii} et~al.(2007){Khmelinskii}, {Lawrence}, {Roostalu}, and
  {Schiebel}]{khmelinskii_cdc14regulated_2007}
{Khmelinskii}, A., C.~{Lawrence}, J.~{Roostalu}, and E.~{Schiebel}, 2007.
\newblock Cdc14-regulated midzone assembly controls anaphase {{B}}.
\newblock \emph{J Cell Biol} 177:981--993.

\bibitem[{Fededa} and {Gerlich}(2012)]{fededa_molecular_2012}
{Fededa}, J.~P., and D.~W. {Gerlich}, 2012.
\newblock Molecular control of animal cell cytokinesis.
\newblock \emph{Nat Cell Biol} 14:440--447.

\bibitem[{Subramanian} et~al.(2013){Subramanian}, {Ti}, {Tan}, {Darst}, and
  {Kapoor}]{subramanian_marking_2013}
{Subramanian}, R., S.-C. {Ti}, L.~{Tan}, S.~A. {Darst}, and T.~M. {Kapoor},
  2013.
\newblock Marking and {{Measuring Single Microtubules}} by {{PRC1}} and
  {{Kinesin}}-4.
\newblock \emph{Cell} 154:377--390.

\bibitem[{Subramanian} et~al.(2010){Subramanian}, {Wilson-Kubalek}, {Arthur},
  {Bick}, {Campbell}, {Darst}, {Milligan}, and
  {Kapoor}]{subramanian_insights_2010}
{Subramanian}, R., E.~M. {Wilson-Kubalek}, C.~P. {Arthur}, M.~J. {Bick}, E.~A.
  {Campbell}, S.~A. {Darst}, R.~A. {Milligan}, and T.~M. {Kapoor}, 2010.
\newblock Insights into {{Antiparallel Microtubule Crosslinking}} by {{PRC1}},
  a {{Conserved Nonmotor Microtubule Binding Protein}}.
\newblock \emph{Cell} 142:433--443.

\bibitem[{Bringmann} et~al.(2004){Bringmann}, {Skiniotis}, {Spilker},
  {Kandels-Lewis}, {Vernos}, and {Surrey}]{bringmann_kinesinlike_2004}
{Bringmann}, H., G.~{Skiniotis}, A.~{Spilker}, S.~{Kandels-Lewis}, I.~{Vernos},
  and T.~{Surrey}, 2004.
\newblock A {{Kinesin}}-like {{Motor Inhibits Microtubule Dynamic
  Instability}}.
\newblock \emph{Science} 303:1519--1522.

\bibitem[{Castoldi} and {Vernos}(2006)]{castoldi_chromokinesin_2006}
{Castoldi}, M., and I.~{Vernos}, 2006.
\newblock Chromokinesin {{Xklp1 Contributes}} to the {{Regulation}} of
  {{Microtubule Density}} and {{Organization}} during {{Spindle Assembly}}.
\newblock \emph{Mol. Biol. Cell} 17:1451--1460.

\bibitem[{Parmeggiani} et~al.(2003){Parmeggiani}, {Franosch}, and
  {Frey}]{parmeggiani_phase_2003}
{Parmeggiani}, A., T.~{Franosch}, and E.~{Frey}, 2003.
\newblock Phase {{Coexistence}} in {{Driven One-Dimensional Transport}}.
\newblock \emph{Phys. Rev. Lett.} 90:086601.

\bibitem[{Juh{\'a}sz}(2007)]{juhasz_weakly_2007}
{Juh{\'a}sz}, R., 2007.
\newblock Weakly coupled, antiparallel, totally asymmetric simple exclusion
  processes.
\newblock \emph{Phys. Rev. E} 76:021117.

\bibitem[{Ashwin} et~al.(2010){Ashwin}, {Lin}, and
  {Steinberg}]{ashwin_queueing_2010}
{Ashwin}, P., C.~{Lin}, and G.~{Steinberg}, 2010.
\newblock Queueing induced by bidirectional motor motion near the end of a
  microtubule.
\newblock \emph{Phys. Rev. E} 82:051907.

\bibitem[{Braun} et~al.(2011){Braun}, {Lansky}, {Fink}, {Ruhnow}, {Diez}, and
  {Janson}]{braun_adaptive_2011}
{Braun}, M., Z.~{Lansky}, G.~{Fink}, F.~{Ruhnow}, S.~{Diez}, and M.~E.
  {Janson}, 2011.
\newblock Adaptive braking by {{Ase1}} prevents overlapping microtubules from
  sliding completely apart.
\newblock \emph{Nat. Cell Biol.} 13:1259--1264.

\bibitem[{Lansky} et~al.(2015){Lansky}, {Braun}, {L{\"u}decke}, {Schlierf},
  {ten~Wolde}, {Janson}, and {Diez}]{lansky_diffusible_2015}
{Lansky}, Z., M.~{Braun}, A.~{L{\"u}decke}, M.~{Schlierf}, P.~R. {ten~Wolde},
  M.~E. {Janson}, and S.~{Diez}, 2015.
\newblock Diffusible {{Crosslinkers Generate Directed Forces}} in {{Microtubule
  Networks}}.
\newblock \emph{Cell} 160:1159--1168.

\bibitem[{Bieling} et~al.(2010{\natexlab{b}}){Bieling}, {Kronja}, and
  {Surrey}]{bieling_microtubule_2010}
{Bieling}, P., I.~{Kronja}, and T.~{Surrey}, 2010.
\newblock Microtubule {{Motility}} on {{Reconstituted Meiotic Chromatin}}.
\newblock \emph{Current Biology} 20:763--769.

\bibitem[{Pierobon} et~al.(2006){Pierobon}, {Mobilia}, {Kouyos}, and
  {Frey}]{pierobon_bottleneckinduced_2006}
{Pierobon}, P., M.~{Mobilia}, R.~{Kouyos}, and E.~{Frey}, 2006.
\newblock Bottleneck-induced transitions in a minimal model for intracellular
  transport.
\newblock \emph{Phys. Rev. E} 74:031906.

\bibitem[{Yildiz} et~al.(2003){Yildiz}, {Forkey}, {McKinney}, {Ha}, {Goldman},
  and {Selvin}]{yildiz_myosin_2003}
{Yildiz}, A., J.~N. {Forkey}, S.~A. {McKinney}, T.~{Ha}, Y.~E. {Goldman}, and
  P.~R. {Selvin}, 2003.
\newblock Myosin {{V Walks Hand-Over-Hand}}: {{Single Fluorophore Imaging}}
  with 1.5-nm {{Localization}}.
\newblock \emph{Science} 300:2061--2065.

\end{thebibliography}

\clearpage
\section*{Figure Legends}

\subsubsection*{Figure~\ref{cartoon}.}
Model and results overview. (A) Schematic of the model of
  motor motion on an antiparallel microtubule overlap. Two filaments
  (green and blue) are modeled as 1D lattices with their plus ends
  oppositely oriented. The filaments are labled R (L) if the plus end
  is pointing to the right (left). Motors (red) bind to empty lattice
  sites with rate $\konc$ and unbind with rate $\koff$. Bound motors
  step toward the MT plus end with rate $v$ (if the adjacent site
  toward the MT plus end is empty) or switch to the other MT with rate
  $s$ (if the corresponding site on the adjacent MT is empty). At MT
  minus ends, motors are inserted at rate $\alpha v$. At MT plus ends,
  motors are removed at rate $\beta v$. (B) Simulated experimental
  images made from our kMC model. Green, motor density. Red, overlap
  region. Scale bar, 5 $\mu$m. Simulations used to generate these
  images used the reference parameter set (table \ref{tab:units}) and
  the indicated bulk motor concentrations. (C) Simulated kymograph
  made from our kMC model with motor spatial position on the
  horizontal axis and time increasing down. Horizontal scale bar, 10
  $\mu$m. Vertical scale bar, 5s. The simulations used to generate the
  kymograph used the reference parameter set and 0.5 nM bulk motor
  concentration.

\subsubsection*{Figure~\ref{fig:phases}.}

Nonequilibrium phases.  Left: motor density $\rho_R (x)$ on MT with
rightward-moving motors. Right: total motor density
$\rho_R(x) + \rho_L (x)$ on both MTs in the overlap.  The parameters
used are the reference parameter set of table \ref{tab:units} with a
bulk motor concentration of 200 nM, except for the switching rate
which is 0.5 \pers\ for the LHLH curve and 0.1 \pers\ for all other
curves and the boundary conditions as noted below. LHLH: low
density-high density-low density-high density coexistence
($\alpha=0.3$, $\beta=0.9$, orange). H$_{\rm e}$: high-density
end-accumulating phase ($\alpha=0.95$, $\beta=0.99$, light blue).
H$_{\rm c}$: high-density center-accumulating phase ($\alpha=0.6$,
$\beta= 0.95$, dark blue). LH$_{\rm c}$: low density-high density
center-accumulating coexistence ($\alpha=0.3$, $\beta=0.95$, green).
M: Meissner phase $\alpha=0.6$, $\beta=0.4$, yellow).  L$_{\rm e}$:
low-density end-accumulating phase ($\alpha=0.3$, $\beta=0.4$, black).
LH$_{\rm e}$: low density-high density end-accumulating coexistence
($\alpha=0.1$, $\beta= 0.95$, magenta).  L$_{\rm c}$: low-density
center-accumulating phase ($\alpha=0.05$, $\beta= 0.1$, red).

\subsubsection*{Figure~\ref{phase_diagram}.}

Phase diagrams. Left: low switching rate (0.1 \pers); right: high
switching rate (0.5 \pers). Solid lines indicate phase boundaries, and
dotted lines boundaries between center-accumulating (v) and
end-accumulating (e) overlap motor density profiles.  The bulk motor
concentration is 200 nM, the motor speed is 5 \mum\pers, and other
parameters are the reference values (table \ref{tab:units}).

\subsubsection*{Figure~\ref{fig:density_x}.}

Motor density profiles for the reference parameter set. Left: motor
density $\rho_R (x)$ on MT with rightward-moving motors. Right: total
motor density $\rho_R(x) + \rho_L (x)$ on both MTs in the
overlap. Parameters are the reference parameter set of table
\ref{tab:units} with the bulk motor concentrations indicated in the
legend.

\subsubsection*{Figure~\ref{fig:center_conc}.}

Motor density at the center of the overlap. Left: variation with bulk
motor concentration; right: variation with motor speed. Points
indicate simulation results and dashed lines theoretical prediction
from Eqn.~\ref{eq:center_dens}. The red line shows the value of the
Langmuir density $\rho_0$ for the same parameters.

\subsubsection*{Figure~\ref{fig:density_speed}.}

Dependence of density profiles on motor speed. Left: motor density
$\rho_R (x)$ on MT with rightward-moving motors. Right: total motor
density $\rho_R(x) + \rho_L (x)$ on both MTs in the overlap.  Varying
motor speed can significantly alter the motor density at the center of
the overlap.  These simulations use the large-$S$ parameter set of
table \ref{tab:units_extended} with bulk motor concentration $c=200$
nM and the motor speeds indicated in the legend.

\begin{figure}[t]
\begin{centering}
\includegraphics[width=0.45 \textwidth]{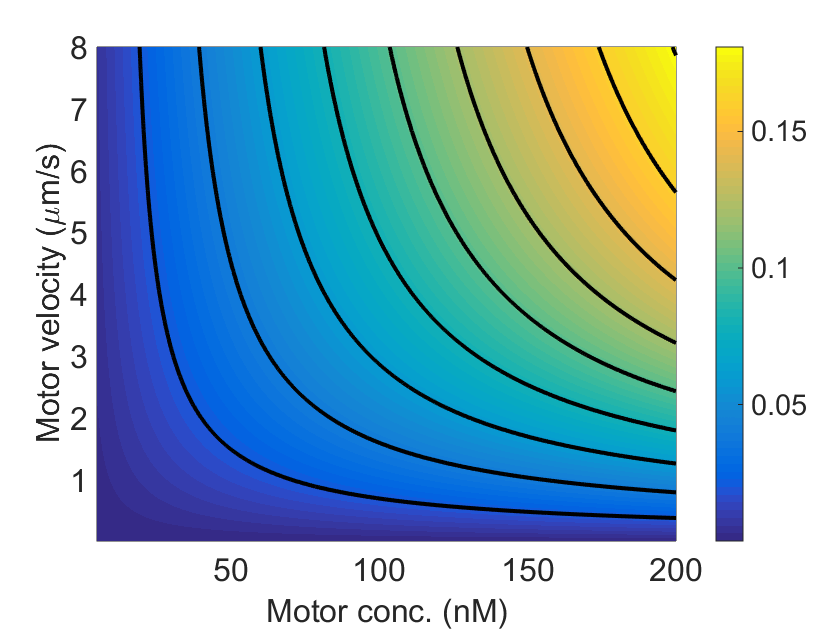}
\includegraphics[width=0.45 \textwidth]{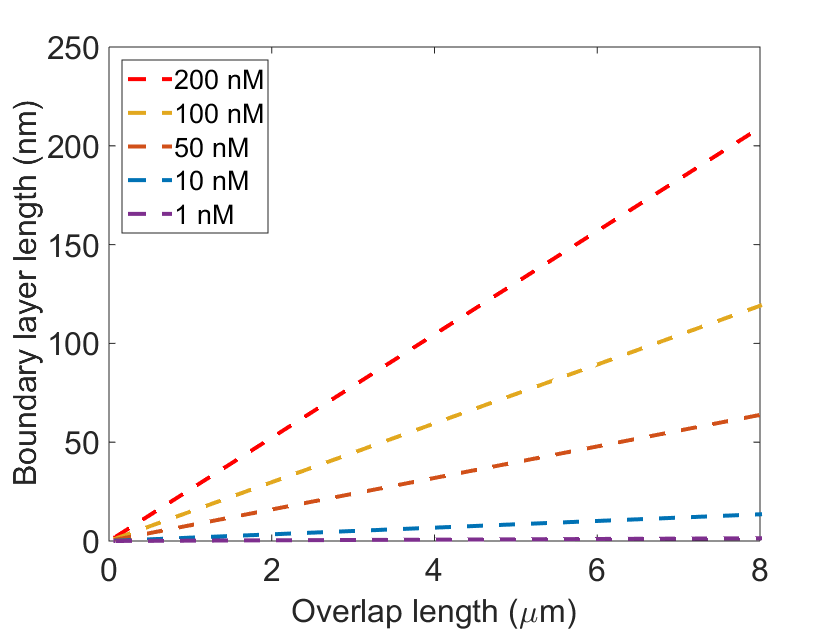}
\caption{Control of boundary layer length.  Left: boundary layer
  length as a fraction of total overlap length, shown as a function of
  bulk motor concentration and motor speed. Right: boundary layer
  length as a function of overlap length for the reference parameter
  set and varying bulk motor concentration. The boundary layer length
  is the distance at each end of the overlap where significant motor
  accumulation occurs and is determined by Eqn.~\ref{eq:xbl}.}
\label{predictions}
\end{centering}
\end{figure}

\subsubsection*{Figure~\ref{predictions}.}

Control of boundary layer length.  Left: boundary layer length as a
fraction of total overlap length, shown as a function of bulk motor
concentration and motor speed. Right: boundary layer length as a
function of overlap length for the reference parameter set and varying
bulk motor concentration. The boundary layer length is the distance at
each end of the overlap where significant motor accumulation occurs
and is determined by Eqn.~\ref{eq:xbl}.

\clearpage

\end{document}